\documentclass[12pt]{article}
\usepackage{epsfig}
\def\lesssim{\mathrel{\mathpalette\vereq<}}

\makeatletter
\def\vereq#1#2{\lower3pt\vbox{\baselineskip1.5pt \lineskip1.5pt
\ialign{$\m@th#1\hfill##\hfil$\crcr#2\crcr\sim\crcr}}}
\makeatother

\title{Can $\varepsilon'/\varepsilon$ be supersymmetric?}

\renewcommand{\thefootnote}{\fnsymbol{footnote}}

\author{Hitoshi Murayama\thanks{Department of Physics, University of
    California, Berkeley, CA 94720, and Lawrence Berkeley National
    Laboratory, MS 50A-5101, Berkeley, CA 94720}}
\date{}
\begin{document}

\begin{titlepage}
\begin{center}
August 23, 1999     \hfill    LBNL-44172 \\
~{} \hfill UCB-PTH-99/32  \\
~{} \hfill hep-ph/9908442\\

\vskip .1in

{\large \bf Can $\varepsilon'/\varepsilon$ be supersymmetric?}%
\footnote{This work was supported in part by the U.S. 
Department of Energy under Contracts DE-AC03-76SF00098, in part by the 
National Science Foundation under grant PHY-95-14797, and in part also 
by Alfred P. Sloan Foundation.}%
\footnote{Invited talk presented at The Chicago Conference on Kaon 
Physics, KAON '99, June 21--26, 1999, Department of Physics, The University 
of Chicago, Chicago, IL}

\vskip 0.3in

Hitoshi Murayama$^{1,2}$

\vskip 0.05in

{$^1$\em Department of Physics, University of California\\
Berkeley, CA 94720-7300}

\vskip 0.05in

and

\vskip 0.05in

{$^2$\em Theoretical Physics Group, MS 50A-5101\\
Lawrence Berkeley National Laboratory, Berkeley, CA 94720}

\end{center}

\vfill

\begin{abstract}
I first motivate why we may want to look at possible new physics
contributions to $\varepsilon'$ given relatively clear experimental but
unclear theoretical situations.  I reexamine the supersymmetric
contribution to $\varepsilon'$ and find an important one generally
missed in the literature.  Based on rather model-independent arguments
based on flavor symmetries, an estimate of the possible supersymmetric
$\varepsilon'$ is given, which interestingly come around the reported
values without fine-tuning.  If the observed values are dominated by
supersymmetry, it is likely to give interesting consequences on
hyperon CP violation, $\mu \rightarrow e\gamma$, and neutron and
electron electric dipole moments.  
    
\end{abstract}

\vfill

\end{titlepage}

\newpage
\setcounter{page}{1}
\setcounter{footnote}{0}
\renewcommand{\thefootnote}{\arabic{footnote}}

\maketitle

\begin{abstract}
I first motivate why we may want to look at possible new physics
contributions to $\varepsilon'$ given relatively clear experimental but
unclear theoretical situations.  I reexamine the supersymmetric
contribution to $\varepsilon'$ and find an important one generally
missed in the literature.  Based on rather model-independent arguments
based on flavor symmetries, an estimate of the possible supersymmetric
$\varepsilon'$ is given, which interestingly come around the reported
values without fine-tuning.  If the observed values are dominated by
supersymmetry, it is likely to give interesting consequences on
hyperon CP violation, $\mu \rightarrow e\gamma$, and neutron and
electron electric dipole moments.  
\end{abstract}

\section{Motivation}

The year 1999 has already seen impressive progress in flavor physics.
CDF reported the first measurement of $\sin 2\beta$ from $B$ decay
\cite{CDF} which strongly hints to CP violation in a system other than
the neutral kaon system.  The situation of direct CP violation
$\varepsilon'$ in the neutral kaon system used to be somewhat unclear, but
KTeV result \cite{KTeV} and the NA48 result reported at this meeting 
\cite{NA48} made the experimental situation basically settled.  The numbers 
on $\varepsilon'/\varepsilon$ reported are\footnote{The world average (W.A.)
  value includes the rescaling of errors due to the PDG prescription
  to account for still-not-too-good $\chi^2/{\rm d.o.f.}=2.8$.  }
\begin{displaymath}
  \begin{array}{ccr}
    {\rm E731} & & (7.4 \pm 5.9) \times 10^{-4} \\
    {\rm NA31} & & (23.0 \pm 6.5) \times 10^{-4} \\
    {\rm KTeV} & & (28.0 \pm 4.1) \times 10^{-4} \\
    {\rm NA48} & & (18.5 \pm 7.3) \times 10^{-4} \\ \hline
    {\rm W.A.} & & (21.2 \pm 4.6) \times 10^{-4} \\
    \end{array}
\end{displaymath}

\begin{table}[th]
\caption{Various standard Model estimates of
    $\varepsilon'/\varepsilon$ in the literature.   Two Bosch'
  estimates use different renormalization schemes (NDR and HV).} 
\label{tab:estimates}
\begin{center}
\begin{tabular}{lc}
Reference & $\varepsilon'/\varepsilon$ \\ \hline
Ciuchini \cite{Ciuchini} & $(4.6 \pm 3.0 \pm 0.4) \cdot 10^{-4}$\\
Bosch (NDR)\cite{Bosch} & $(7.7^{+6.0}_{-3.5}) \cdot 10^{-4}$ \\
Bosch (HV) \cite{Bosch} & $(5.2^{+4.6}_{-2.7}) \cdot 10^{-4}$ \\
Bertolini \cite{Bertolini} & $(17 ^{+14}_{-10}) \cdot 10^{-4}$ \\
\end{tabular}
\end{center}
\end{table}

\begin{figure}
	\epsfxsize=0.7\textwidth
	\centerline{\epsfbox{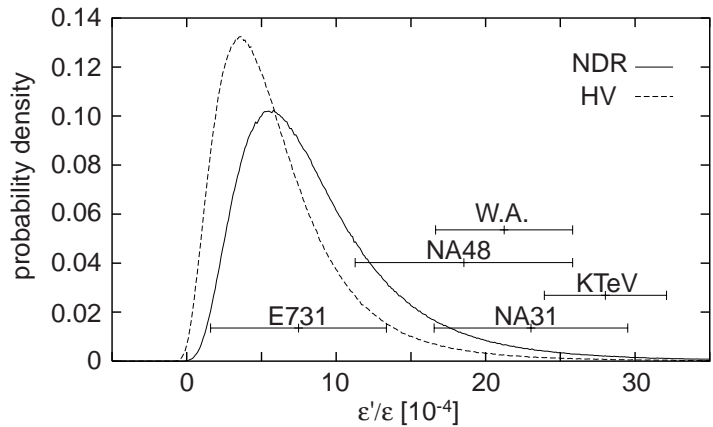}}
\caption{Probability density distributions for the standard model 
  calculations of $\varepsilon'/\varepsilon$ in NDR and HV schemes from 
  \cite{Bosch} compared to experimental values.}
\label{fig:PDF}
\end{figure}

On the other hand, the theoretical situation is rather unclear.  The
calculation of $\varepsilon'$ in the Standard Model is difficult partly
because of a cancellation between gluon and electroweak
penguins which makes the result sensitive to the precise values of the
hadronic matrix elements.  A (not complete) list of
theoretical calculations is given in Table~\ref{tab:estimates}.
See \cite{Buras} and \cite{panel} for more details on this issue.  The
experimental values are compared to the ``probability density
distributions for $\varepsilon'/\varepsilon$'' \cite{Bosch} in
Fig.~\ref{fig:PDF}.  There is a feeling in the community that the
data came out rather high, even though one cannot draw a definite
conclusion if the Standard Model accounts for the observed high value
because of theoretical uncertainties.

Of course the correct strategy to resolve this issue is to improve
theoretical calculations, probably relying on the progress in lattice
calculations.  Unfortunately, this is a challenging program and we
cannot expect an immediate resolution.  What I instead attempt in this
talk is the alternative approach: think about new physics candidates
which are ``reasonable'' and at the same time account for the observed
value of $\varepsilon'/\varepsilon$.  Since any new physics explanation
would probably give rise to other consequences, the cross-check can
start eliminating such possibilities.  If all ``reasonable''
candidates get excluded, one can draw a conclusion that the Standard
Model should account for the observed high value of $\varepsilon'$.  Or
such a cross-check might confirm other consequences which would be
truly exciting.  

Naturally, I turned my attention to supersymmetry, most widely
discussed candidate of physics beyond the Standard Model, and asked
the question if supersymmetric contribution to $\varepsilon'$ can be
interesting at all.

\section{Lore}

There are many studies of $\varepsilon'$ in supersymmetric
models, most notably \cite{GG}.  Their detailed study found
the possible range of supersymmetric contribution,
$(\varepsilon'/\varepsilon)_{\rm SUSY} = (-0.4$--$7.4)\times 10^{-4}$.  Then
clearly supersymmetry cannot account for the observed high value, and
indeed I do not have anything new to add to their beautiful analysis
within the framework they used: the minimal supergravity model.

The reason why supersymmetric contribution to $\varepsilon'$ is small
can be quite easily understood.  Within the minimal supergravity
framework, the main contribution to flavor-changing effects between
the first and second generations originate in the left-handed squark
mass-squared matrix.  In general, the superpartners of the left-handed
quarks have a mass-squared matrix
\begin{equation}
  {\cal L}_{LL} = - \left( \begin{array}{cc}
        \tilde{d}_L^* & \tilde{s}_L^*
  \end{array}
   \right)
  \left( \begin{array}{cc}
      m_{11}^2 & m_{12}^2 \\ m_{12}^{2*} & m_{22}^2
    \end{array}
  \right)
  \left( \begin{array}{cc}
      \tilde{d}_L \\ \tilde{s}_L
    \end{array}
  \right) .
  \label{eq:LL}
\end{equation}
If there is a non-vanishing off-diagonal element $m_{12}^2$ in the
above mass-squared matrix, it would contribute to flavor-changing
processes from loop diagrams.  The easiest way to study such
contributions is to use the mass insertion formalism where one treats
the off-diagonal element perturbatively as an ``interaction'' in the
squark propagator, because the existent constraints require the
off-diagonal element to be small anyway.  The size of such a
perturbation can be nicely parameterized by $(\delta_{12}^d)_{LL}
\equiv m_{12}^{2}/m^2$ where $m^2$ is the average of the two diagonal
elements.

\begin{figure}[thbp]
  \begin{center}
    \leavevmode
	\epsfxsize=0.45\textwidth
	\epsfbox{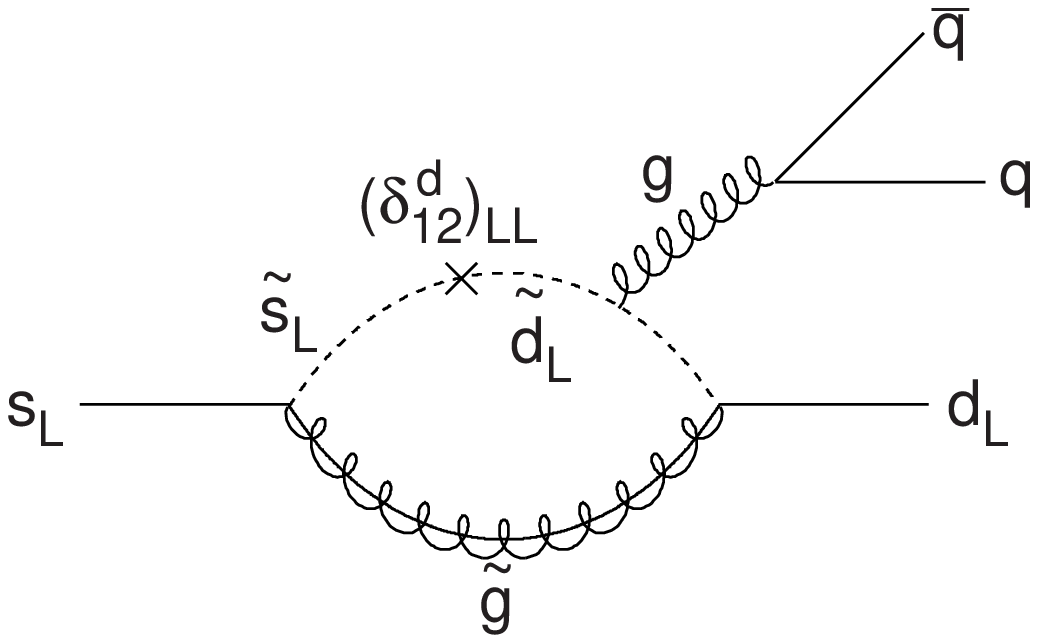} \qquad
	\epsfxsize=0.45\textwidth
	\epsfbox{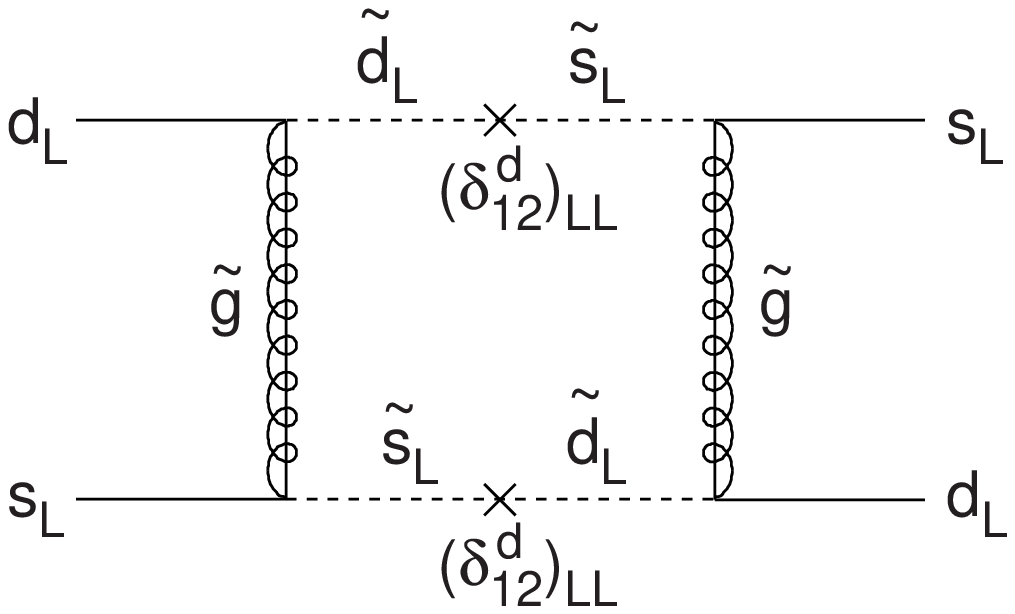}
    \caption{Representative Feynman diagrams for LL contributions to
    $\varepsilon'$ (left) and $\Delta m_K$, $\varepsilon$ (right).} 
    \label{fig:LL}
  \end{center}
\end{figure}

Contribution of the mass insertion to $\varepsilon'$ is given by%
\footnote{Here and hereafter, I take $M_{\tilde{g}} = m$ for
  simplicity.  See \cite{Gabbiani} for more details.} \cite{Gabbiani}:
\begin{equation}
  \left(\frac{\varepsilon'}{\varepsilon}\right)_{\rm SUSY} 
  \simeq 27 \times 10^{-4} 
  \left( \frac{500 \rm GeV}{m} \right)^2
  \frac{{\rm Im} (\delta_{12}^d)_{LL}}{0.50} .
\end{equation}
A representative Feynman diagram is shown in Fig.~\ref{fig:LL}.
On the other hand, the insertion of the same parameter
$(\delta_{12}^d)_{LL}$ induces kaon mixing parameters 
\begin{eqnarray}
  (\Delta m_K)_{\rm SUSY} & \simeq & (\Delta m_K)_{\rm experiment}
  \left( \frac{500 \rm GeV}{m} \right)^2
  \frac{{\rm Re} (\delta_{12}^d)^2_{LL}}{(0.04)^2} \\
  \varepsilon_{\rm SUSY} & \simeq & \varepsilon_{\rm experiment}
  \left( \frac{500 \rm GeV}{m} \right)^2
  \frac{{\rm Im} (\delta_{12}^d)^2_{LL}}{(0.003)^2} .
\end{eqnarray}
A representative Feynman diagram is shown in Fig.~\ref{fig:LL}.
Therefore, once the constraints from $\Delta m_K$ and $\varepsilon$ are
satisfied, it does not allow a large contribution to $\varepsilon'$
unless some fine-tuning is done.  This simple comparison already
gives a typical order of magnitude $(\varepsilon'/\varepsilon)_{\rm SUSY}
\lesssim 2 \times 10^{-4}$, with some wiggle room by varying
$M_{\tilde{g}}$, allowing chargino contributions, etc.\footnote{One 
can maximize $\varepsilon' \propto {\rm Im} (\delta_{12}^d)_{LL}$ 
beyond this estimate
by assuming that $(\delta_{12}^d)_{LL}$ is nearly pure imaginary 
because $(\delta_{12}^d)^2_{LL}$ is then nearly pure real and hence 
$\varepsilon$ constraint is satisfied.} This 
leads to the basic conclusion in \cite{GG} that the supersymmetric 
contribution to $\varepsilon'$ is never important.\footnote{This 
conclusion, however, relied only on $\Delta I =1/2$ amplitude from 
supersymmetry.  A generally missed contribution to $\Delta I =3/2$ 
amplitude due to the 
isospin breaking $m^{2}_{\tilde{d}_{R}} \neq m^{2}_{\tilde{u}_{R}}$
leads to a too-large $\varepsilon'$ if the splitting is $O(1)$.  
A more natural size of splitting 
from the renormalization-group point of view is of order 10\%, and with 
a quarter of the maximum possible value of ${\rm Im} (\delta_{12}^d)_{LL}$, the
resulting size of $\varepsilon'$ is in the ballpark of the observed 
value \cite{KN}.}

\section{New Contribution}

Antonio Masiero and myself found that there is a loophole in previous 
discussions why supersymmetric contribution to $\varepsilon'$ is small 
\cite{MM}.\footnote{Another possibility of generating $\varepsilon'$ 
from an enhanced $\overline{d_{L}}\gamma^{\mu}s_{L} Z_{\mu}$ vertex was 
suggested \cite{CI} which is subject to tighter constraints 
phenomenologically \cite{BCIRS}.} We pointed out that a broad class of 
supersymmetric models actually gives an important 
contribution,\footnote{This estimate is larger by a factor of two than 
that in \cite{MM}, which ignored the running effect of the dipole 
moment operator.}
\begin{equation}
  \left( \frac{\varepsilon'}{\varepsilon} \right)_{\rm SUSY} 
  \simeq 30 \times 10^{-4} 
  \left( \frac{500 \rm GeV}{m} \right)^2 \frac{\sin \phi}{0.5}
  \times (0.5\mbox{--}2) \times (0.5\mbox{--}2)
  \label{eq:epsilonp}
\end{equation}
where the uncertainties come from model dependence and hadronic
matrix elements.  I discuss the angle $\phi$ and the origin of this
new contribution below.

The masses of quarks arise from the coupling of left-handed and
right-handed quarks to the Higgs expectation value,
\begin{equation}
  {\cal L}_{\rm quark\ mass} = 
  - \left( \begin{array}{cc}
        \overline{d_R} & \overline{s_R}
  \end{array}
  \right)
  \left( \begin{array}{cc}
      y_{11} & y_{12} \\ y_{21} & y_{22}
    \end{array}
  \right)
  \left( \begin{array}{cc}
      d_L \\ s_L
    \end{array}
  \right)
  \langle H_{d} \rangle + c.c.
\end{equation}
Here, $H_d$ is Higgs field which couples to the down-type quarks in
the Minimal Supersymmetric Standard Model.  Similarly, in addition to
the mass-squared mass matrix (\ref{eq:LL}) for the left-handed
squarks and the analogous one for the right-handed squarks, there is
another contribution to the masses of squarks which mix left-handed
and right-handed particles, 
\begin{equation}
  {\cal L}_{LR} = 
  - \left( \begin{array}{cc}
        \tilde{d}_R^* & \tilde{s}_R^*
  \end{array}
  \right)
  \left( \begin{array}{cc}
      A_{11} & A_{12} \\ A_{21} & A_{22}
    \end{array}
  \right)
  \left( \begin{array}{cc}
      \tilde{d}_L \\ \tilde{s}_L
    \end{array}
  \right)
  \langle H_{d} \rangle + c.c.
  \label{eq:LR}
\end{equation}
The important point here is, without going into detailed discussions
of models, we expect the matrix $A_{ij}$ to have similar structure as
the Yukawa matrix $y_{ij}$ because they have exactly the same property
under the flavor symmetries.  I will describe this point more clearly
in the context of models later on, but based on this simple fact, I
expect
\begin{equation}
  A_{ij} = m_{\rm SUSY} Y_{ij} \times O(1),
\end{equation}
where $m_{\rm SUSY}$ is the typical mass scale of supersymmetry breaking
({\it i.e.}\/, superparticle masses).  It is possible that $A_{ij}$
and $Y_{ij}$ are exactly proportional to each other as matrices, which
was assumed in the minimal supergravity and hence in previous
analyses, but not necessarily.  Generically, one expects the
following patterns for Yukawa and $A$ matrices,
\begin{eqnarray}
  & &\left( \begin{array}{cc}
      y_{11} & y_{12} \\ y_{21} & y_{22}
    \end{array}
  \right) \langle H_d \rangle = 
  \left( \begin{array}{cc}
      m_d & m_s \lambda \\  & m_s
    \end{array}
  \right) ,\\
  & &\left( \begin{array}{cc}
      A_{11} & A_{12} \\ A_{21} & A_{22}
    \end{array}
  \right) \langle H_d \rangle = 
  m_{\rm SUSY} \left( \begin{array}{cc}
      c m_d & b m_s \lambda \\  & a m_s
    \end{array}
  \right) + c.c. \, ,
  \label{eq:A}
\end{eqnarray}
where $a$, $b$, $c$ are $O(1)$ unknown constants and $\lambda = \sin
\theta_{C}$.  The (2,1) elements are intentionally left blank because
we do not know the mixing angles of right-handed quarks (read:
model-dependent).

\begin{figure}[t]
  \begin{center}
    \leavevmode
	\epsfxsize=0.45\textwidth
	\epsfbox{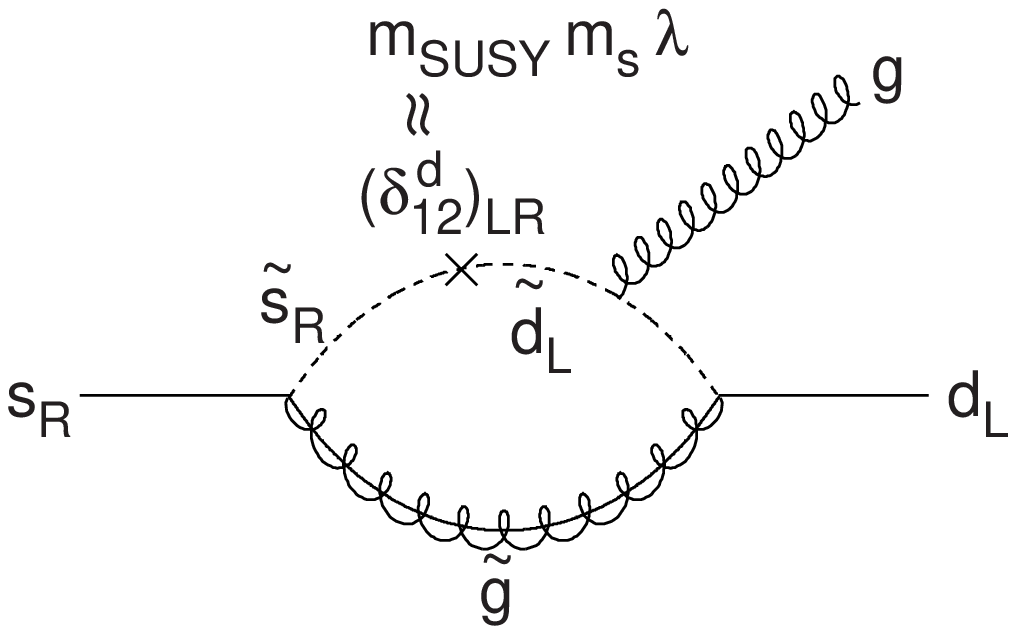} \qquad
	\epsfxsize=0.45\textwidth
    \epsfbox{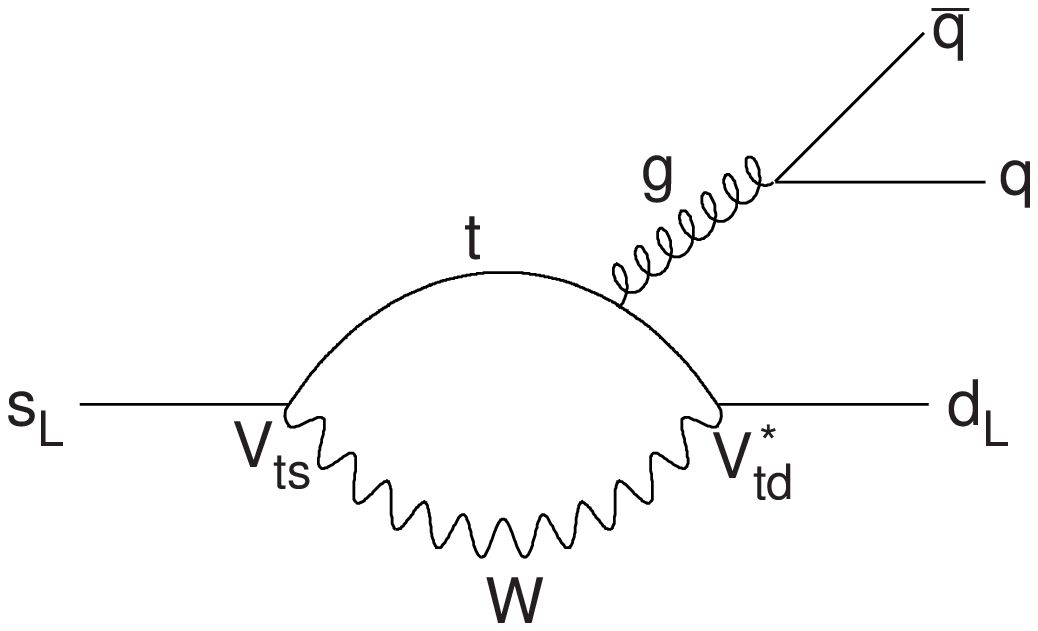}
    \caption{Representative Feynman diagrams for the supersymmetric
      contribution from LR mass insertion (left), and the standard model
      contribution (right) to $\varepsilon'$.}
    \label{fig:LR}
  \end{center}
\end{figure}

In order to discuss flavor-changing effects from squarks, one first
needs to go to the mass basis for quarks, which is achieved by
rotating the Yukawa matrix by the Cabibbo angle $\theta_C$.  The same
rotation acts on the $A$ matrix, but unless $a=b$, it still leaves the
off-diagonal element after rotation $A'_{12} \langle H_d \rangle =
(b-a) m_{\rm SUSY} m_s \lambda$.  We can again describe its effect in
terms of a mass insertion parameter
\begin{equation}
  (\delta_{12}^d)_{LR} = \frac{A'_{12} \langle H_d \rangle}{m^2} 
  = 4 \times 10^{-5} \left( \frac{m_s}{100~{\rm MeV}} \right)
  \left(\frac{m_{\rm SUSY}}{m}\right) \left(\frac{\rm 500~GeV}{m}\right) 
  \times (b-a).
\end{equation}
It contributes to the flavor-changing chromo-electric dipole moment 
$\frac{g_s}{8\pi^2} m_s$ $\overline{d_L} \sigma^{\mu\nu} G_{\mu\nu} s_R$ 
(Fig.~\ref{fig:LR}), where $G_{\mu\nu}$ is the gluon field strength.  
This operator generates a supersymmetric contribution to 
$\varepsilon'$ as $(\varepsilon'/\varepsilon)_{\rm SUSY} = 30 \times 
10^{-4} ({\rm Im} (\delta_{12}^d)_{LR}/ 2 \times 10^{-5})$, which gives 
Eq.~(\ref{eq:epsilonp}).  The model uncertainty (0.5--2) comes from 
the lack of knowledge on the $O(1)$ constant $a,b$, and also that we 
do not know the ratio of $m_{\rm SUSY}$ in the LR mass matrix relative to 
the average squark mass.  The phase factor is $\phi = {\rm arg} 
(b-a)$.  Finally the matrix element of this operator between kaon and 
two-pion states vanishes at the lowest order in chiral perturbation 
theory, and it was estimated in \cite{BEF} using the chiral quark 
model.  I have arbitrarily assigned a factor of two uncertainty 
(0.5--2) as a guess.\footnote{The authors of \cite{BCIRS} instead 
assign an uncertainty of $B_G = (1$--$4)$ in their notation and have 
emphasized that the sign of the matrix element is model-dependent.} 

Note that the exact proportionality between Yukawa and $A$ matrices 
assumed in the minimal supergravity automatically makes the 
off-diagonal element vanish after diagonalizing the Yukawa matrix and 
hence no flavor-changing effects.  This is one of the reasons why the 
LR mass insertion has been regarded unimportant in the literature.  
But such an exact proportionality is a strong assumption with no 
justification.  The constraint from $\varepsilon$ on LR mass insertion 
is an order of magnitude weaker: ${\rm Im}(\delta_{12}^d)_{LR}^2 
\lesssim (3.5 \times 10^{-4})^2 (m^2/{\rm 500~GeV})^2$ 
\cite{Gabbiani}.  This phenomenological fact that the constraint from 
$\varepsilon$ still allows an interesting LR contribution to 
$\varepsilon'$ has been known (see, {\it e.g.}\/, \cite{Gabbiani}).  
But the natural size of $(\delta_{12}^d)_{LR}$ has not been discussed 
and it was always assumed to be smaller than the size estimated here 
due to various prejudices partly because of the constraint from the 
neutron electric dipole moment (see discussions later).

The reason why this supersymmetric contribution is competitive (or 
larger) than the standard model contribution is basically because of 
the larger KM factor as seen in the following back-of-envelope 
estimate.  The estimate of the supersymmetry contribution is given by 
$\displaystyle \frac{g_s^3}{8\pi^2} \frac{m_s \lambda}{m_{\rm SUSY}^2} 
\overline{d_L} \sigma^{\mu\nu} G_{\mu\nu} s_R$ and is suppressed only 
by the Cabibbo angle in the LR mass insertion.  On the other hand the 
standard model QCD penguin is suppressed because all three generations 
should participate for CP-violation: $\displaystyle \frac{g_s^2 
g^2}{8\pi^2} \frac{V_{ts} V_{td}^*}{m_t^2} \overline{d_L} \gamma^\mu 
T^a s_L $ $\displaystyle\overline{q} \gamma_\mu T^a q$.  To compare 
them, $m_s$ in the supersymmetric contribution can be regarded as an 
$O(1)$ parameter in kaon decays.  The mass scale is higher $m_{\rm SUSY} > 
m_t$, but the KM factor is larger by 100.  However the matrix element 
of the chromo-electric dipole moment operator vanishes at the leading 
order in chiral perturbation theory \cite{BEF}, resulting in about a 
factor of ten suppression, which brings the supersymmetry contribution 
roughly on par with the QCD penguin.  Finally, the QCD model penguin 
is partially canceled by the electroweak penguin.  This leaves the 
supersymmetry contribution important in $\varepsilon'$.

\section{Models}

Now that we have found the new supersymmetric contribution to
$\varepsilon'$ based on general grounds, it is important to ask what
class of models actually gives such a contribution.  Because we are
talking about flavor physics issues in supersymmetry here (see
\cite{Hall} for a good summary in this context), we need to look at
models which naturally explain the hierarchy in fermion masses.

The easiest example to explain is the model based on U(2) flavor
symmetry \cite{U(2)}.  In this model, the first and second generations
form a doublet under the U(2) flavor symmetry, and the third
generation a singlet.  This makes the LL squark mass-squared matrix
(\ref{eq:LL}) an identity matrix and helps to keep the first and
second generation scalars to avoid too-large flavor-changing effects
in $\Delta m_K$ and $\varepsilon$.  The flavor symmetry allows the top
Yukawa coupling, but forbids down and strange Yukawa couplings and
hence needs to be broken by some vacuum expectation values (VEVs).
They introduce two such VEVs, one in the symmetric tensor of U(2)
$\sigma_{ij} = \sigma_{ji}$ and the other in the anti-symmetric tensor
$\alpha_{ij} = -\alpha_{ji}$, which can generate the Yukawa matrix
\begin{equation}
  \frac{1}{M} \left( \begin{array}{cc}
      \overline{d_R} & \overline{s_R} 
    \end{array} 
  \right)
  \left( \begin{array}{cc}
      0 & \langle \alpha_{12}\rangle \\ 
      -\langle\alpha_{12}\rangle & \langle \sigma_{22} \rangle
    \end{array}
  \right)
  \left( \begin{array}{cc}
      d_L \\ s_L
    \end{array}
  \right) H_{d} ,
\end{equation}
where $M$ is the energy scale of the flavor physics.  The point here
is that (2,2) and (1,2) elements originate from couplings to different
fields $\sigma$ and $\alpha$ and hence their supersymmetry breaking
counterparts can have different coefficients, giving in general $a
\neq b$ in the LR mass matrix (\ref{eq:A}).

A similar situation occurs in the string theory, where the Yukawa
couplings $y_{ij}$ are in general functions of so-called moduli fields
$T$: $y_{ij}(T)$.  The moduli fields also acquire supersymmetry
breaking VEV ($F_T$), and the LR mass matrix is given by $A_{ij} =
(\partial y_{ij}(T)/\partial T) F_T$.  Generally, these two matrices
are not proportional to each other, even though their structures are
likely to be similar if the Yukawa hierarchy is natural, {\it i.e.}\/,
stable under small changes of the moduli.  A particular realization of
this idea is an attempt to generate Yukawa hierarchy as a consequence
of a somewhat large compactification radius of an orbifold and fields
of different generations belonging to twisted sectors of different
fixed points.  Then the modular invariance requires Yukawa couplings
to be proportional to powers of Dedekind eta function $\eta(T)$:
$Y_{ij} \simeq \eta(T)^{3+n_i+n_j+n_H}$ ($n_i$ is the modular weight)
\cite{Ibanez}.  This mechanism can generate a Yukawa hierarchy because
$\eta (T) \sim e^{-T}$ for a large $T$.  Then the derivative picks the
power $3+n_i+n_j+n_H$ which is generation dependent and hence $b\neq
a$ again.  The flavor-changing effects from the LL mass insertion, 
however, is a problem in general string models.

I have to emphasize that not all supersymmetric models give this type 
of contributions to $\varepsilon'$.  For instance, the models with 
gauge-mediated supersymmetry breaking (see \cite{GR} for a review) 
give either vanishing $A$-matrix or one proportional to the Yukawa 
matrix and hence does not give interesting $\varepsilon'$.  If the 
Cabibbo angle originates in the up sector, the naive analysis here 
also fails.  Also, the $A$-matrix may be real.  All I can argue here 
is that we generically expect supersymmetric contribution to 
$\varepsilon'$ as in (\ref{eq:epsilonp}), but there are exceptions.  
See \cite{models} for more discussions on model-dependence.

\section{Other Observables}

Since supersymmetry can give a contribution to $\varepsilon'$ or an
interesting size, we should now ask if there are other consequences of
this.  The immediate question is the neutron electric dipole moment.
Again using the mass insertion technique, the current constraint is
$|{\rm Im}(\delta_{11}^d)_{LR} | \lesssim 3 \times 10^{-6}$
\cite{Gabbiani}, while my estimate is $(\delta_{11}^d)_{LR} \equiv
A_{11} \langle H_d \rangle / m^2 \simeq m_{\rm SUSY} m_d/m^2 \simeq 1
\times 10^{-5} (500~{\rm GeV}/m_{\rm SUSY})$.  If we are interested in a
large supersymmetric contribution to $\varepsilon'$, it is natural to
expect a phase of order unity in $A_{11}$ as well.  This basically
saturates the current upper limit for a phase of order a third, and an
improve experimental bound would be extremely interesting.  The phase
in $A_{11}$, however, is in principle unrelated to that in $A_{12}$,
and in fact some models give a real $A_{11}$ despite a complex
$A_{12}$ \cite{BDM}.

A more exciting aspect is that hyperon CP violation arise from the
same chromo-electric dipole moment operator which generates
supersymmetric $\varepsilon'$ and in principle there is a perfect
correlation.  Unfortunately the matrix element uncertainties obscures
the correlation but Pakvasa argued that the hyperon CP violation
should appear within the sensitivity of the current HyperCP
experiment if $\varepsilon'$ is saturated by supersymmetry in the way
I described \cite{Pakvasa}.  It would the most direct test of this
possibility.  

Another important question is what the corresponding flavor-changing
effects in the lepton sector would be.  In fact, we found a new
contribution to $\mu \rightarrow e \gamma$ which was also generally
missed in the literature.  Unfortunately the estimate is uncertain
because we do not know the mixing angle in the lepton sector, but a
reasonable guess is $V_{\nu_e \mu} \simeq 0.016$--0.05 suggested by
the small angle MSW solution.  The new limit from MEGA collaboration
\cite{MEGA} ${\rm BR}(\mu \rightarrow e\gamma) < 1.2 \times 10^{-11}$
(90\% CL) leads to the bound on the mass insertion parameter
$(\delta_{12}^l)_{LR} < 4.2 \times 10^{-6} (m_{\tilde{l}}/{\rm 500
  GeV})$.  This is to be compared to the estimate
$|(\delta_{12}^l)_{LR}| \simeq m_{\rm SUSY} m_\mu V_{\nu_e \mu} /
m_{\tilde{l}}^2 \simeq 6.6 \times 10^{-6} ({\rm 500
  GeV}/m_{\tilde{l}}) (V_{\nu_e \mu}/0.032) (m_{\rm SUSY}/m_{\tilde{l}})$.
Note that this bound is on the {\it magnitude}\/ of the parameter, not
its imaginary part, and is hence more difficult to avoid.  One can
regard this estimate as a lower bound on $m_{\tilde{l}}$ of about
500~GeV, which is much stronger than previously discussed limits.  The
electron electric dipole moment $d_e = (0.18 \pm 0.16) \times
10^{26} < 0.38 \times 10^{-26} e{\rm cm}$ (90\% CL) translates to the
limit $|{\rm Im}(\delta_{11}^l)_{LR}| < 1.0 \times 10^{-6}
(m_{\tilde{l}}/{\rm 500 GeV})$, compared to the estimate
$(\delta_{11}^l)_{LR} \simeq 1.0 \times 10^{-6} (m_{\tilde{l}}/{\rm
  500 GeV})$.  The supersymmetric contribution again nearly saturates
the constraint and a future improvement of the bound would be useful.

More recently, it has been suggested that there may also be 
enhancements in $K_{L} \rightarrow \pi^{0} e^{+} e^{-}$ \cite{BCIRS} 
and $K \rightarrow \pi \pi \gamma$ \cite{CIP}.  See also discussions 
in \cite{others}.

Overall, the naive estimate of the LR mass insertion I've described
gives interesting and important contributions to $\varepsilon'$,
$\mu \rightarrow e\gamma$, $d_n$, and $d_e$ and those on the first two
had been generally missed in the literature.

\section{Conclusions}

We reexamined the supersymmetric contribution to $\varepsilon'$, and
found that an important contribution had been missed in the
literature.  The assumptions that went into the estimate are: (a)
hierarchical Yukawa matrix is probably controlled by certain flavor
symmetries, (b) $A$-matrix is subject to the same control
which makes its structure similar to the Yukawa matrix but not
necessarily exactly proportional to each other, (c) Cabibbo rotation
arises in the down sector, and (d) $O(1)$-ish phase in the $A$-matrix.
Given these assumptions, we came up with the estimate
Eq.~(\ref{eq:epsilonp}) which surprisingly is in the ballpark of
reported experimental values without any fine-tuning.  

If $\varepsilon'$ is dominated by the supersymmetric contribution
described here, it should give important contribution to the hyperon
CP violation and possibly also to the neutron electric dipole moment.
Correspondingly, there is an important contribution to $\mu \rightarrow
e\gamma$ and possibly also to the electron electric dipole moment.

I thank Antonio Masiero for collaboration.  This work was supported in
part by the U.S. Department of Energy under Contracts
DE-AC03-76SF00098, in part by the National Science Foundation under
grant PHY-95-14797, and in part by Alfred P. Sloan Foundation.

\end{document}